\documentclass[twocolumn,nofootinbib,amsmath,amssymb,a4paper]{revtex4}

\usepackage{hyperref}
\usepackage{graphicx}
\usepackage{dcolumn}
\usepackage{bm}

\newcommand{\bmath}{\begin{displaymath}}
\newcommand{\emath}{\end{displaymath}}
\newcommand{\KL}{$\rm K_{L}$}
\newcommand{\klpi}{$\rm K_{2\pi}$}
\newcommand{\KS}{$\rm K_{S}$}
\newcommand{\Kp}{$\rm K^{+}$}
\newcommand{\Km}{$\rm K^{-}$}
\newcommand{\Kpm}{$\rm K^{\pm}$}
\newcommand{\ke}{$\rm K_{e3}$}
\newcommand{\km}{$\rm K_{\mu3}$}
\newcommand{\kche}{$\rm K^{\pm}_{e3}$}
\newcommand{\kchm}{$\rm K^{\pm}_{\mu3}$}
\newcommand{\kchpi}{$\rm K^{\pm}_{2\pi}$}
\newcommand{\zi}{$\Xi^{0}$~}
\newcommand{\zib}{$\Xi^{0} \rightarrow~\Sigma^{+}~e^{-}~\bar\nu_{e}$~}
\newcommand{\vus}{$\vert V_{us} \vert$}

\newcommand{\lp}{$\lambda_{+}$}
\newcommand{\lz}{$\lambda_{0}$}

\def\Journal#1#2#3#4{{#1} {\bf #2}, #3 (#4)}

\def\PLB{{\em Phys. Lett.}  B}
\def\PRL{\em Phys. Rev. Lett.}
\def\PRD{{\em Phys. Rev.} D}
\def\ZPC{{\em Z. Phys.} C}
\def\EPJ{{\em Eur. Phys. J.} C}
\def\JPG{\em Journ. Phys. G}

\begin{document}

\title{NA48 Results on Kaon and Hyperon Decays Relevant to \vus}

\author{M. Veltri}
 \email{veltri@fis.uniurb.it}
\affiliation{
Istituto di Fisica, Universit\`a di Urbino and INFN Sez. Firenze\\
Via S. Chiara 27, I--61029 Urbino, Italy}

\begin{abstract}
\noindent New results from the NA48 experiment on kaon and hyperon decays relevant
to \vus~ are reported here. For charged kaons we present measurements of 
$BR(K^{\pm} \to \pi^{0} e^{\pm} \nu) $ and 
$BR(K^{\pm} \to \pi^{0} \mu^{\pm} \nu) $.
On neutral kaon decays we report the measurements of BR($K_{L} \to \pi^+ \pi^-$)
and of $K_{L} \rightarrow \pi^{\pm} \mu^{\mp} \nu$ form factors slopes. 
For hyperons we present results on the BR($\Xi^{0} \rightarrow~\Sigma^{+}~e^{-}~\bar\nu_{e}$).
\end{abstract}

\maketitle

\section{Introduction}
The NA48 experiment at the CERN SPS has been taking data from 1997 to 2004.
During its first phase (ended in 2001) employed simultaneous \KL/\KS~ beams and was
devoted to the precision measurement of direct CP violation in the neutral 
kaon system. Also several semileptonic and rare \KL~ decays could be studied.\\
The second phase (NA48/1) was active in the year 2002. Using only the \KS~
beam at an increased intensity the experiment measured rare \KS~ and neutral
hyperon decays.\\
In the third phase (NA48/2) the beam set--up was changed again in order to 
simultaneously collect \Kp~ and \Km~ decays. The data taking was in progress
during the years 2003--2004 with the purpose to search for direct CP violation
in the \Kpm~ decays to three pions together with the high statistics measurements 
of semileptonic and rare charged kaon decays.\\
A detailed description of the NA48 detector can be found elsewhere~\cite{na48_det}.
The most relevant components are a high resolution liquid--krypton electromagnetic 
calorimeter (LKr) and a magnetic spectrometer consisting of 4 drift chambers and 
a dipole magnet located inside a helium tank. Other components are a hodoscope
for precise track time determination, a hadronic calorimeter and a muon
counter (MUC). Reported here are the results from kaon and hyperon decays relevant
to the determination of \vus.

\subsection{$K_{\ell 3}$ decays}
The semileptonic kaon decays ($K_{\ell 3}, \ell=e, \mu$), being pure vector 
transitions, provide the most accurate and theoretically cleanest way to 
measure \vus. The matrix element of the decay can be written in terms of two 
dimensionless form factors $f_{\pm}(t)$:
\begin{eqnarray}
 {\cal M} = \frac{G_{F}}{\sqrt{2}} V_{us}
 \left[ f_{+}(t) \left(P_{K}+P_{\pi}\right)^{\mu} 
  + f_{-}(t) \left(P_{K}-P_{\pi}\right)^{\mu} \right] \nonumber \\
  \times~ \bar u_{\ell} \gamma_{\mu} (1+\gamma_{5}) u_{\nu}~~~~~~
\end{eqnarray}
where $G_{F}$ is the Fermi coupling constant,
$P_{K/\pi}$ are the kaon/pion four--momenta, $u_{\ell, \nu}$ the 
lepton spinors and $t=(P_{K}-P_{\pi})^2$. 
Being proportional to the lepton mass squared, 
the contribution of $f_{-}$ can be neglected in $K_{e3}$ decays.
The $K_{\mu3}$ decays are usually described in terms of the scalar form factor
defined as $f_{0}(t) = f_{+}(t) + t/(m^{2}_{K} - m^{2}_{\pi}) f_{-}(t)$;
$f_{+}$ and $f_{ 0}$ are related to the vector ($1^{-}$) and scalar 
($0^{+}$) exchange to the lepton system, respectively.
The master formula for the $K_{\ell 3}$ decay rates (fully inclusive of 
radiation) is:
\begin{eqnarray}
    \Gamma(K_{\ell 3}) =  
     \frac{C^{2}_{K} G^{2}_{F} m^{5}_{K}}{192 \pi^{3}}
     S_{EW} \vert V_{us} \vert^{2} \vert f_{+}(0) \vert^{2} \nonumber \\
     I^{\ell}_{K} \left[ f_+(t),f_0(t) \right]
    (1 +  \delta^{\ell K}_{SU(2)} + \delta^{\ell K}_{EM}) 
  \label{eq:kl_rate}
\end{eqnarray}
\noindent where the index $K$ stands for $ K^{\pm}, K^{0}$, 
$ m_{K}$ is the appropriate kaon mass, $C^{2}$ is 1 for $K^{0}$ and 1/2 
for $K^{\pm}$ decays, $S_{EW}$ is the short--distance electroweak correction,
$f_{+}(0)$ is the calculated form factor at zero momentum transfer.
For vector transitions \cite{AdemolloGatto} the SU(3) breaking appears
only to second order, as consequence $f_{+}(0)$ is close to unity. This term 
is calculated for $K^{0}$ transitions. The quantity $\delta^{\ell K}_{SU(2)}$,
(which is 0 for $K^{0}$ decays) is common for $e$ and $\mu$ channels and accounts 
for the SU(2) breaking corrections to  $f_{+}(0)$ in $ K^{\pm}$ decays. 
The $\delta^{\ell K}_{EM}$ term represents the long--distance
radiative corrections, these are important only for the $K^{0}$ case and especially
for the $\mu$ channel. $I^{\ell}_{K} \left[ f_+(t),f_0(t) \right]$ is the phase
space integral which depends on the two form factors which describe the decays.

\section{The branching ratio of \kche~and \kchm~decays}
NA48/2 has measured~\cite{na48_BRKch} the branching ratios of 
$K^{\pm} \to \pi^{0}   e^{\pm} \nu $ (\kche) and 
$K^{\pm} \to \pi^{0} \mu^{\pm} \nu $ (\kchm) decays studying data 
collected in a dedicated minimum bias run. 
As normalization channel the decay 
$K^{\pm} \to \pi^{\pm}   \pi^{0} $ (\kchpi) was used. 
The basic selection of all the three modes was common and was based on the presence 
of only one track in the spectrometer and at least two clusters (photons) in the 
LKr that were consistent with a $\pi^{0}$ decay. Further kinematical and particle 
id requirements were applied to separate the three decays. The invariant mass for
\kchpi~candidates was required to be within three sigma of the reconstructed 
kaon mass, that is: 472.2 MeV/c$^2 < m_{\pi^{\pm} \pi^{0}} <$ 510.2 MeV/c$^2$ while 
for $K^{\pm}_{\ell 3}$ candidates it had to be outside this range. 
In order to separate two from three body decays cuts were applied to the 
squared missing mass ($m_{\nu}^{2}$), assuming the decaying kaon to have an 
energy of 60 GeV and direction along the beam line axis.
To distinguish electrons from pions a cut was imposed on the ratio E/p, between the 
energy deposited by the track in the LKr and its momentum measured by the spectrometer. 
\kche~(\kchpi) events were identified  requiring this ratio to be greater (smaller) 
than 0.95. The identificaton of \kchm~ events instead was based on the association, 
in space and time, of the extrapolation  of the track and a hit in the MUC. 
The selected samples were practically background free and amounted to about 
87000 \kche, 77000 \kchm~and  718000 \kchpi~events.
After correcting for acceptance, trigger and particle id efficiences,
radiative effects and taking the current PDG value for the \kchpi~ branching 
fraction~\cite{PDG06} we obtained (results $\times 10^{-2}$):
\begin{eqnarray}
BR(K^{\pm}_{e3})   = 5.221\pm0.019_{stat}\pm0.008_{syst}\pm0.030_{norm} \\
BR(K^{\pm}_{\mu3}) = 3.425\pm0.013_{stat}\pm0.006_{syst}\pm0.020_{norm} 
\end{eqnarray}
The uncertainty is dominated by the existing data for the BR(\kchpi).
Higher branching fractions with respect to PDG ones are found for both
\kche~ and \kchm~ confirming the results reported by the BNL--E865 
collaboration~\cite{BNL-865}.\\
Using the newly measured BRs we determined, through Eq.~\ref{eq:kl_rate}
(see~\cite{na48_BRKch} for details on the values used for the phase space 
integrals and radiative corrections), the following values for the product
of the \vus~ matrix element and $f_+(0)$:
\vskip -0.2cm
\begin{table}[h]
\begin{tabular}{rclr}
$\vert V_{us} \vert f_{+}(0)$ & = & $0.2204\pm0.0012$ & $\left[ K^{\pm}_{e3} \right]$ \\
                              & = & $0.2177\pm0.0013$ & $\left[ K^{\pm}_{\mu3} \right] $ 
\end{tabular}
\end{table}

Combining these values by assuming $\mu-e$ universality, we obtained:
\begin{equation}
\vert V_{us} \vert f_{+}(0) = 0.2197\pm0.0012~~~~~~~\left[ K^{\pm}_{\ell3} \right]
\end{equation}
We can compare the above result to the prediction obtained by imposing the
unitarity condition on the CKM matrix. Assuming that 
$f_{+}(0)=0.961\pm0.008$ ~\cite{leutwyler-roos} and using the latest values
of $\vert V_{ud} \vert$ and $\vert V_{ub} \vert$ from ~\cite{PDG06} we
have $\vert V_{us} \vert_{\mathrm unitarity} f_{+}(0) = 0.2185\pm0.0022$ in good
agreement with our result.

\section{The branching ratio of $K_{L} \rightarrow \pi^+ \pi^-$ decay}
\label{section_kl2pi}
This measurement~\cite{na48_BRK2pi} of the branching ratio of the 
$K_{L} \rightarrow \pi^+ \pi^-$ (\klpi) decay is based on data taken during a 
dedicated two--days run in 1999 by the NA48 experiment. 
Throughout this run only the \KL~beam was present. 
The event acquisition was triggered by the presence of two particles in the 
charged hodoscope system and of a vertex of two tracks in the spectrometer. 
As normalization channel the 
$K_{L} \rightarrow \pi^{\pm} e^{\mp} \nu_{e}$ ($K_{e3}$)
decay was used. The majority of the event selection was common to both
decay modes requiring the tracks to have opposite charge and imposing
cuts for the definiton of the vertex. To select \klpi~events a cut on
the two track invariant mass was applied. The abundant semileptonic 
background was removed by requiring both tracks to have E/p$<$0.93 
(against $e$) and 
rejecting events with an associated signal in the muon counter ($K_{\mu3}$
events with $\pi$ decay). 
The $K_{e3}$ events used for normalization, being the only relevant 
\KL~ decay channel with an electron in the final state, 
were selected by only further demanding a track with E/p$>$0.93.
\begin{figure}[h]
\vskip -0.35 cm
\includegraphics[width=0.47\textwidth]{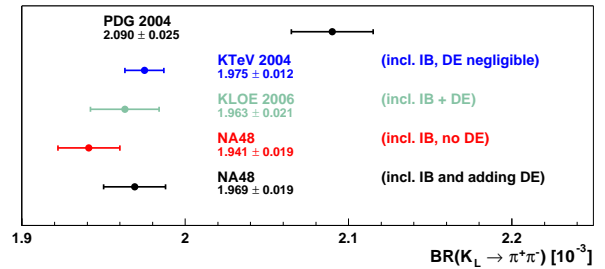}
\vskip -0.35 cm
\caption{\label{fig:BR_KL2pi} Comparison of results for $BR(K_L \to \pi^+ \pi^-)$.}
\end{figure}
The final samples had a statistics of about 47000 \klpi~and five million $K_{e3}$ 
events. Since the $\pi^{+} \pi^{-}$ event selection did not imply any requirement
for or against photons, the radiative $K_L \to \pi^+ \pi^- \gamma$ decays were
also accepted. To calculate the branching ratio we chose to exclude the CP conserving
contribution from radiative decays with the $\gamma$ coming from direct emission (DE),
considering only the CP violating contribution from inner bremsstrahlung (IB).
Using the MC the DE contribution was determined and subtracted from the signal.
To extract BR(\klpi) the value of BR($K_{e3}$)=0.4022$\pm$0.0031 was used.
This value is the NA48 measurement~\cite{na48_BRKe3} updated following the improved 
knowledge of $BR(K_L \to 3 \pi^0)$. 
Finally we obtained:
\begin{equation}
BR(K_L \to \pi^+ \pi^- + \pi^+ \pi^- \gamma_{IB}) = 
(1.942 \pm 0.019)\times 10^{-3}
\end{equation}
Our result is in good agreement with the measurements done by 
KTeV~\cite{ktev_BRK2pi} and KLOE~\cite{kloe_BRK2pi}. 
All results contradict the values reported by PDG 2004~\cite{PDG04}.

\section{ $K_L \to \pi^{\pm} \mu^{\mp} \nu~(K_{\mu3})$ form factors}
\begin{figure}[h]
\includegraphics[width=0.4\textwidth]{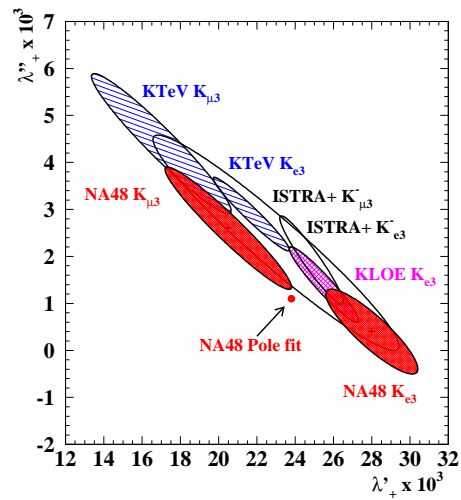}
\caption{\label{fig:ellipse} 1 $\sigma$ contour plots in the plane 
$\lambda^{'}_{+}$--$\lambda^{''}_{+}$ showing the NA48 results together 
with those of \cite{kloe_ff,ktev_ff,na48_ke3ff,istra-e,istra-m} for 
the quadratic fits of the \km~and~\ke~decays.}
\end{figure}

As shown in eq.~\ref{eq:kl_rate} the form factors of the $K_{\ell 3}$
decays are an input (through the phase space integrals) for the evaluation
of \vus. It is customary to expand the form factors up to a linear or a quadratic
term in $t$:
\begin{equation}
f_{+,0}(t) =  f_{+}(0) ~
   \left[ \rule{0mm}{5mm} 1 + \lambda^{'}_{+,0}~ t/m_{\pi}^{2} +
   \frac{1}{2}~ \lambda^{''}_{+,0}~ (t/m_{\pi}^{2})^{2} \right].
\end{equation}
According to the pole model instead, the $t$ dependence can be related to the 
exchange of $K^*$ resonances:
\begin{equation}
f_{+,0}(t) =  f_{+}(0) ~\frac{m^{2}_{V,S}}{m^{2}_{V,S}-t}
\end{equation}
Recently new parametrizations, based on dispersion techniques, have been
proposed~\cite{stern}:
\begin{eqnarray}
f_{+}(t) = f_{+}(0)\exp\Bigl{[}\frac{t}{m^{2}_{\pi}}(\Lambda_{+} + H(t))\Bigr{]}, \\
f_{0}(t) = f_{+}(0)\exp\Bigl{[}\frac{t}{(m_{K}^{2} -  m_{\pi}^{2})}(\ln C - G(t))\Bigr{]}. \nonumber
\end{eqnarray}
The parameter
$\ln C = \ln[ f_{0}(m_{K}^{2} -  m_{\pi}^{2}) ]$ is the logarithm of
the value of the scalar form factor at the Callan--Treiman point.
This value can be used to test the existence of right handed quark currents (RHCs)
coupled to the standard W boson.\\
Using the same data sample utilized for the \klpi~analysis (Sec.~\ref{section_kl2pi})
the $K_{\mu3}$ events were selected~\cite{na48_kmu3ff} by demanding a vertex of two 
tracks of opposite charge. The muon track was identified by a coinciding hit
in the MUC. The $K_{e3}$ background was rejected imposing E/p$<$0.9 to both tracks
and $K_{3\pi}$ background was suppressed applying a cut on the kinematical variable
$P_{0}^{'2}$. The undetected neutrino in $K_{\ell3}$ decays is responsible
for the quadratic ambiguity in the determination of the \KL~energy. A cut was 
imposed requiring that the two kaon energies (called low and high) had to 
be greater than 70 GeV.
Finally a cut was applied on the variable $p_{\nu}^{*} - p_{\nu T}$,
where $p_{\nu}^{*}$ is the total neutrino momentum in the kaon CMS.
This quantity, clearly positive for good \km~events, is highly
sensitive to resolution effects which give rise to a moderate negative
tail. We set a cut at $p_{\nu}^{*} - p_{\nu T}  > 7$ MeV/c, selecting
a region where the MC simulation accurately reproduces the data behaviour.
The final data sample consisted of about 2.3$\times 10^6$ \km~ events.
The determination of the form factor parameters is done studying the Dalitz plot
density. The low energy solution was used to evaluate the muon and pion 
energies (in the kaon center of mass) needed to build the Dalitz plot.
According to the MC simulation this corresponds to the most probable solution,
being in 61\% of cases the correct one. The Dalitz plot was divided
into cells with a dimension of about $4 \times 4$ MeV$^2$.
With this choice, about 39\% of the events are reconstructed exactly in the same 
cell where they were generated.
To extract the form factors we fit the data Dalitz plot corrected
for acceptance, migration of events due to the wrong choice of the kaon energy, 
backgrounds and radiative effects. 
The acceptance and the radiative corrections depend slightly on the form factor
values used in the simulation. In order to reduce possible biases due to this 
dependence, the MC samples were generated, after an iterative procedure, with 
form factor values close to the fitted ones. 
Various $t$ dependences of the form factors were considered:
linear, quadratic, pole and dispersive. The fit results are listed in
Table~\ref{table:fit_results}.
\begin{table}[h]
\caption{ Form factors fit results for linear, quadratic pole and dispersive
parametrizations. The first error is the statistical one, the second the systematic one.}
\label{table:fit_results}
\begin{ruledtabular}
\begin{tabular}{cccc}
\rule{0mm}{3mm}Linear ($\times 10^{-3}$) &  &  & \\
     \lp             &           \lz        & & $\chi^2$/ndf \\
26.7$\pm$0.6$\pm$0.8 & 11.7$\pm$0.7$\pm$1.0 & &  604.0/582 \\ \hline

\rule{0mm}{3mm}Quadratic ($\times 10^{-3}$) & & & \\
  $\lambda^{'}_{+}$    & $\lambda^{''}_{+}$  & \lz                 & $\chi^2$/ndf \\
 20.5$\pm$2.2$\pm$2.4  & 2.6$\pm$0.9$\pm$1.0 & 9.5$\pm$1.1$\pm$0.8 &  595.9/581 \\ \hline

\rule{0mm}{3mm}Pole (MeV/c$^2$) & & & \\
  $m_V$             &      $m_S$         &  & $\chi^2$/ndf \\
 905$\pm$9$\pm$17   & 1400$\pm$46$\pm$53 &  & 596.7/582\\ \hline

\rule{0mm}{3mm}Dispersive ($\times 10^{-3}$) &  &   & \\
$\Lambda_{+}$          & $\ln C$                & & $\chi^2$/ndf \\
 23.3$\pm$0.5$\pm$0.8  & 143.8$\pm$8.0$\pm$11.2 & &  595.0/582 \\
\end{tabular}
\end{ruledtabular}
\end{table}
Our results indicate the presence of a quadratic term in the
expansion of the vector form factor in agreement with other
recent analyses of kaon semileptonic decays.
Fig.~\ref{fig:ellipse} shows the comparison between the results
of the quadratic fits as reported by the recent experiments
\cite{kloe_ff,ktev_ff,na48_ke3ff,istra-e,istra-m}. The 1 $\sigma$~contour plots
are shown, both for \ke~and \km~decays; the ISTRA+ results have been
multiplied by the ratio $(m_{\pi^{+}}/m_{\pi^{0}})^{2}$.
The results are higly correlated, those from this measurement and from
KTeV have a larger quadratic term and appear only in partial agreement
with the other \ke~experiments.
While the result for \lp~ is well compatible with the recent (and most
precise) KTeV measurement, the value of \lz~appears to be shifted towards
lower values.
According to the model proposed in~\cite{stern} the value of $\ln C$
can be used to test the existence of RHCs by comparing it with the Standard
Model predictions.  We obtain for a combination of the RHCs couplings
and the Callan--Treiman discrepancy ($\tilde\Delta_{CT}$) the value:
$2 (\epsilon_S - \epsilon_{NS}) + \tilde\Delta_{CT} =
-0.071 \pm0.014_{NA48} \pm0.002_{theo} \pm0.005_{ext}$,
where the first error is the combination in quadrature of the
to the uncertainties related to the approximations used to replace
the dispersion integrals and the last one is due to the external
experimental input.

\section{The branching ratio of $\Xi^{0}~ \beta$--decay}
In analogy with kaons, the semileptonic \zi $\beta$--decay (\zib) 
can provide a determination of \vus~ via the measurements of BR and lifetimes.
Furthermore this decay allows a test of SU(3) breaking effects. In exact SU(3)
symmetry approximation the form factors describing the \zi $\beta$--decay
are the same of the well known neutron $\beta$--decay.
The NA48/1 experiment collected a sample of 6316 \zib 
events~\cite{na48_BRXi0}.
The identification of the decay was performed using the subsequent 
$\Sigma^{+} \to p \pi^{0}$ decay with $\pi^{0} \to \gamma \gamma$. 
\begin{figure}[h]
\includegraphics[width=0.4\textwidth]{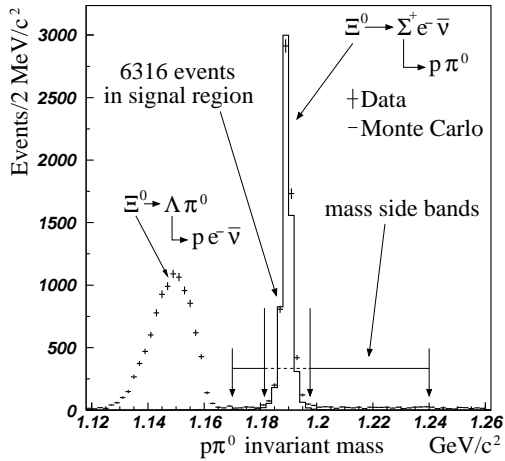}
\caption{\label{fig:SigmaInvMass} Reconstructed $p \pi^0$ invariant mass
distribution far all $\Sigma^{+} \to p \pi^{0}$ candidates after all
selection criteria were applied. The solid line shows the MC prediciton for
the signal.}
\end{figure}
The final state consisted of a proton and  an electron, leaving tracks in the 
spectrometer, in addition to two photons being detected as clusters in the LKr 
calorimeter, and one unobserved anti--neutrino. The event signature was given
by the presence of a $\Sigma^{+}$ since the $\Xi^{0}~ \beta$--decay is the 
only source of of this kind of particles in the neutral beam given that the 
two--body decay ($\Xi^{0} \to \Sigma^+ \pi^-$) is forbidden by energy conservation.
Thus signal events were identified by requiring an invariant $p \pi^{0}$ 
mass consistent with the nominal $\Sigma^{+}$ mass value.
With a background of about 3.4\% and using as normalization channel the
$\Xi^{0} \to \Lambda \pi^{0} $ decay, a value for the branching ratio
was determined: 
\begin{equation}
BR(\Xi^0 \to \Sigma^+ e^- \overline{\nu}_e) =
(2.51 \pm 0.03_{stat}\pm 0.09_{syst})\times10^{-4}~
\end{equation}
where the systematic error is dominated by the trigger efficiency determination,
the geometrical acceptance and the experimental knowledge of the form factors. 
This value is consistent with the KTeV measurement based on 625 events~\cite{ktev_BRXi0}.
Since the trigger did not distinguish between particle charges with respect to
the event hypotheses, the recorded data sample also contained decays of anti--hyperons
allowing the first measurement of the $\overline{\Xi^0}\to\overline{\Sigma^+}e^+ \nu_e$ 
branching ratio to be performed. A sample of 555 candidates was found in the signal
region with a background of 136$\pm$8 events allowing the determination of:
\begin{equation}
BR(\overline{\Xi^0}\to\overline{\Sigma^+}e^+ \nu_e) =
(2.55 \pm 0.14_{stat}\pm{0.10}_{syst}) \times 10^{-4} 
\end{equation}
Using the combined result of the measured $\Xi^0$ and $\bar \Xi^0$ branching ratios
together with the current values of form factors and neglecting SU(3) breaking corrections
we evaluated $\vert V_{us} \vert = 0.209^{+0.023}_{-0.028}$ consistent with the present
value obtained from kaon semileptonic decays~\cite{PDG06}. 
The uncertainty on \vus~ is largely dominated by the error on the form factors. 
Alternatively, the form factors ratio $g_1/f_1$ could 
be extracted using the current \vus~ value determined from kaon decays obtaining:
$ g_{1}/f_{1}=1.20\pm 0.04_{BR} \pm 0.03_{ext}$, where the last uncertainty includes
contributions from \vus, $\Xi^{0}$ lifetime and form factors. The agreement with the 
prediction for exact SU(3) symmetry ($f_1=1.0$; $g_1=1.27$) favours SU(3) breaking 
models that do not modify significantly $g_{1}/f_{1}$.


\end{document}